\documentclass[twocolumn,showpacs,preprintnumbers,amsmath,amssymb,superscriptaddress,prb]{revtex4}

\usepackage{graphicx}
\usepackage{dcolumn}
\usepackage{bm}
\usepackage{subfigure}
\usepackage{ulem}

\begin{document}

\preprint{APS/123-QED}

\title{Terahertz magneto-transport measurements\\in underdoped Pr$_{2-x}$Ce$_x$CuO$_4$ and comparison with ARPES}
\author{G. S. Jenkins}
\author{D. C. Schmadel}%
\author{P. L. Bach}%
\author{R. L. Greene}%
\affiliation{%
Center for Nanophysics and Advanced Materials\\Department of
Physics, University of Maryland, College Park, Maryland 20742, USA
}%

\author{X.~B\'echamp-Lagani\`{e}re}
\author{G. Roberge}
\author{P. Fournier}
\affiliation{Regroupement qu\'eb\'ecois sur les mat\'eriaux de pointe, D\'epartement de Physique, Universit\'e de Sherbrooke, 
Sherbrooke, Qu\'ebec, Canada, J1K 2R1}

\author{H. D. Drew}
\affiliation{%
Center for Nanophysics and Advanced Materials\\Department of
Physics, University of Maryland, College Park, Maryland 20742, USA
}%

\date{\today}

\begin{abstract}
We present magneto-transport measurements performed on underdoped Pr$_{2-x}$Ce$_x$CuO$_4$ at THz frequencies as a function of temperature and doping.  A rapidly decreasing Hall mass is observed as the doping is reduced consistent with the formation of small electron Fermi pockets.  However, both dc and infrared (IR) magneto-transport data strongly deviate from the predictions of transport theory in the relaxation time approximation (RTA) based on angular resolved photoemission data. The Hall mass is observed to increase continuously with increasing temperature with no signature at the N\'eel temperature. In the paramagnetic state, the temperature dependence of the Hall mass is consistent with current vertex corrections to the Hall conductivity due to magnetic fluctuations as observed in overdoped Pr$_{2-x}$Ce$_x$CuO$_4$. Possible causal mechanisms for the discrepancy between transport theory within the RTA and the magneto-transport data are discussed.

\end{abstract}

\maketitle

\section{Introduction}

The cuprates remain an intriguing problem because they are strongly interacting electron systems close to the Mott state characterized by strong competing fluctuations of charge, magnetic, and superconducting order.  The subtle interplay between these fluctuations manifests as different behavior in the various cuprate versions.  Charge order is emphasized near 1/8 doping.  Superconducting fluctuations extend to high temperatures in the hole doped (p-type) cuprates.  Magnetic order is more strongly expressed in the electron (n-type) doped cuprates.  Mott correlations as half filling is approached become particularly important in underdoped materials. The difficulty of describing the electronic properties of the cuprates may lie in the necessity to account for these different fluctuating orders simultaneously.  

The predominance of magnetic order in the n-type cuprates has lead to the suggestion that they approach the Mott state through a spin density wave (SDW) transition at a quantum critical point near optimal doping.\cite{Lin_Millis} Evidence for this SDW order in the magnetic state is seen in the direct observation of a gap in angular resolved photoemission (ARPES) measurements,\cite{ARPES_edoped_Park} and an onset of gap-like optical absorption in the infrared whose frequency increases as half filling is approached.\cite{Lin_Millis,Zimmers_OpticalGap}  No such gap-like feature is present in the p-type cuprates, which instead exhibit a pseudogap corresponding to a suppression of spectral weight near the Fermi level at temperatures below a characteristic pseudogap temperature.\cite{PseudoGapReview} Both the gap and pseudogap produce Fermi arcs in ARPES measurements, wherein portions of the large hole-like Fermi surface (FS) observed at optimal doping are obliterated.\cite{ARPES_FermiArcsReview} In the case of n-type cuprates the obliterated sections are reasonably hypothesized to result from a SDW gaping of the FS from the ($\pi$, $\pi$) magnetic order which doubles the unit cell and reconstructs the FS by zone folding, leading to small electron-like Fermi pockets centered at ($\pi$, 0).\cite{Lin_Millis}

In this paper we examine the dc and infrared (IR) magneto-transport of underdoped n-type cuprates to test the SDW picture by comparing the transport data with ARPES data within conventional transport theory. In the magnetically ordered state with the absence of magnetic fluctuations it might be expected that transport theory within the relaxation time approximation (RTA) would be applicable.  ARPES data provides the FS geometry, the Fermi velocities and the quasiparticle relaxation rates which are the required ingredients for calculating the magneto-conductivity tensor in the RTA. We have measured the Hall angle of four underdoped Pr$_{2-x}$Ce$_x$CuO$_4$ (PCCO) samples at an optical frequency of 10.5 meV. Our principle observation is large discrepancies between the magneto-transport data and the predictions of the RTA based on the Fermi surface properties measured by ARPES.

\section{Measurement and Results}

Thin c-axis oriented films of PCCO were grown via pulsed laser deposition onto LaSrGaO$_4$ (001) substrates. We report data on two samples which were previously measured in the near-infrared \cite{AZimmersFilmGrowth} along with two new similarly grown samples.\cite{PatrickFilmGrowth} The four samples have a chemical doping of x=0.10, 0.12, 0.135, and 0.15, thicknesses of 275, 126, 220, and 145 nm, and Tc's of less than 2K, 2K, 12.5K, and 19.6K, respectively.

\begin{figure}[t]
\includegraphics[scale=.45,clip=true, trim = 50 0 180 0]{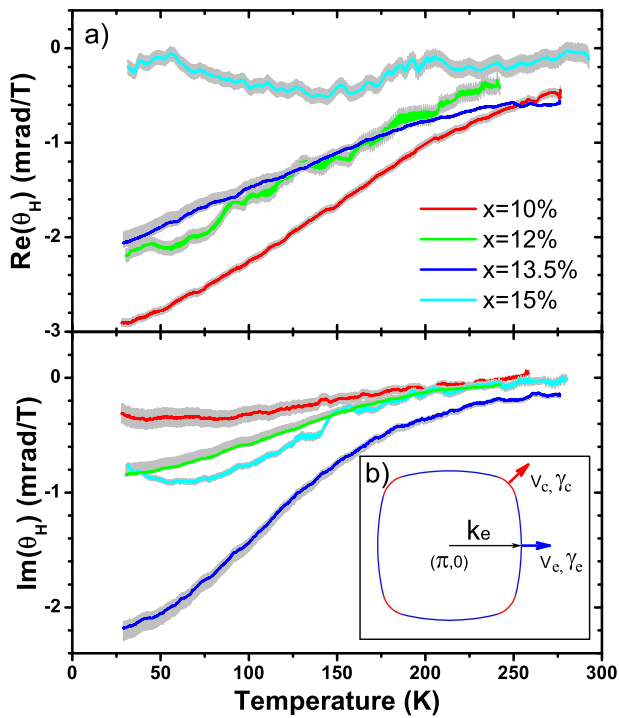}
\caption{\label{fig:ReImHA}
(a) Real and imaginary part of the Hall angle measured at 10.5 meV as a function of temperature. A negative value implies a dominant electron-like response. Grey depicts error bars of one standard deviation. 
(b) Schematic of the FS model of the electron
pocket centered at ($\pi$,0). The blue and red arcs have
different radii of curvature whose center defines the 'edge' and 'corner' of the pocket, respectively.
The distance from the center of the pocket to the edge is $k_e$. The
radius of curvature of the edge arc divided by $k_e$ defines variable
$\beta_e$. $\kappa$ is the fraction of the FS which is made up of
the edge arc lengths. Along with the condition that the tangent to
the FS is continuous at all points, the three variables $k_e$, $\beta_e$, and $\kappa$
define the size and shape of the FS. The corner and edge arc are assigned Fermi velocities and scattering rates as measured by ARPES.}
\end{figure}

The Faraday rotation and circular dichroism, characterized by the complex Faraday angle $\theta_F$, were measured at 10.5 meV as
a continuous function of temperature. The output of a far-infrared
molecular vapor laser was polarization modulated with a rotating
quartz 1/4-wave plate and subsequently transmitted through the
sample at normal incidence. Magnetic fields up to 8T were applied parallel to the c-axis of the film. The detector signal was
harmonically analyzed and the complex Faraday
angle extracted, a technique that is detailed elsewhere.\cite{GJThesis}

In the thin film limit, the complex Hall angle is derived from the
Faraday angle via $\theta_H = (1+ \frac{n+1}{Z_0 \sigma_{xx} d})
\theta_F$, where $n$ is the index of refraction of the substrate,
$Z_0$ is the impedance of free space, and $d$ is the thickness of
the film. The film thickness was chosen to allow
sufficient transmitted power while maintaining a relatively small
conversion factor. FTIR spectroscopic transmission measurements
were performed in the spectral range from 2 to 13 meV at a set of
discrete temperatures ranging from 10 to 300 K. The complex
conductivity was extracted by fitting the transmission data to a simple Drude model.\cite{RefFit}

The Hall frequency $\omega_H $
and Hall scattering rate $\gamma_H$ are defined in terms of
the complex Hall angle within a Drude parameterization:
\begin{eqnarray}
\theta_H = \frac{\sigma_{xy}}{\sigma_{xx}} =
\frac{\omega_H}{(\gamma_H - \it{i}\omega)}\label{eq:HAdefinition}
\end{eqnarray}
where $\sigma_{xy}$ is the Hall conductivity, $\sigma_{xx}$ is the
longitudinal conductivity, $\omega_H=q B / (m_H c)$ is the Hall
frequency,  $\omega$ is the radiation frequency, $m_H$ is the
effective Hall mass, B is the applied magnetic field, and q is the
effective charge of the quasiparticle. The Hall mass can
be expressed as:
\begin{eqnarray}
\frac{m_H}{m_e}  = -\frac{\omega_c^e}{\omega}
Im(1/\theta_H)\label{eq:mHFromHall}
\end{eqnarray}
where $\omega_c^e = 0.115 meV / T$ is the bare electron cyclotron
frequency.

Various calibrations of the IR Hall system were performed to ensure the accuracy and validity of the measurement technique.\cite{GJThesis} A polarizer and waveplate substituted for the sample induce a rotation and ellipticity which are easily calculated and compared to measurements. To fully test the system under similar operating conditions as the measurements performed on high-Tc cuprates, a GaAs 2-DEG heterostructure was measured as a function of magnetic field with a thin film of NiCr deposited on the surface with a 25\,$\Omega$ sheet resistance to lower the transmitted power to a level commensurate with a typical high-Tc sample. The cyclotron resonance of a 2-DEG produces a resonance and antiresonance associated with the imaginary and real part of the Faraday angle, respectively. Since the resonance and antiresonance are simply related, the relative sizes of the real and imaginary Faraday angle allow careful comparison with measurements. In addition, measurements performed on optimally doped single crystal Bi$_{2}$Sr$_{2}$CaCu$_{2}$O$_{8+x}$ measured at 10.5, 5.25, and 3.0 meV when analyzed with Eq.\,\ref{eq:mHFromHall} demonstrate a temperature and frequency independent Hall mass\cite{GJThesis} consistent with ARPES,\cite{ARPES_Valla} the optical mass derived from ac conductivity,\cite{Tu} and the Hall mass derived from near-infrared Hall measurements.\cite{DSThesis}

\begin{figure}[t]
\includegraphics[scale=.38,clip=true, trim = 70 30 90 60]{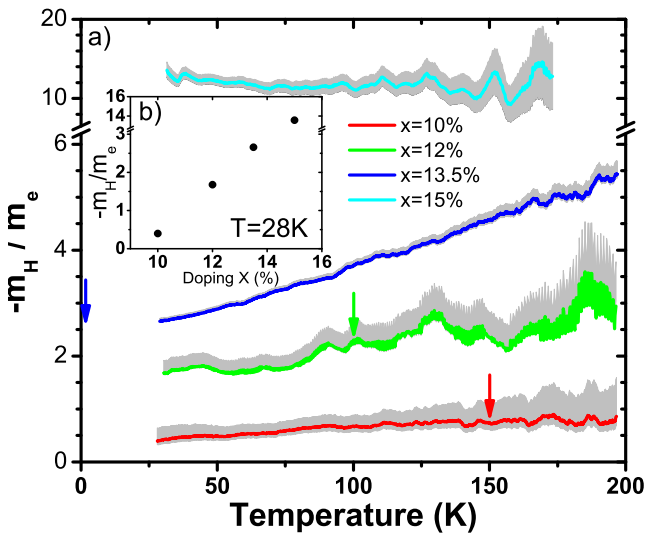}
\caption{\label{fig:mHandModelDiagram}
(a) The Hall mass extracted from Fig.\,\ref{fig:ReImHA}a using Eq.\,\ref{eq:mHFromHall} normalized to the bare electron mass. Temperatures above 200 K are excluded due to noise introduced into $m_H$ from the exceedingly small Hall angle. The negative value of the Hall mass signifies a net electron-like response.  The extremely small Hall masses associated with the 10\% and 12\% samples exemplify a large IR Hall response which cannot be obtained from ARPES data analyzed in terms of the RTA. Failure of the system to recover to a large hole-like Fermi surface (positive $m_H$) well above the N\'eel temperature (marked by arrows)\cite{Motoyama} indicates that current vertex corrections induced by antiferromagnetic fluctuations in the paramagnetic state are active. (b) The low temperature (28\,K) Hall mass plotted versus doping.}
\end{figure}

For all PCCO samples, the measured complex Faraday angle at fixed temperature was linear in applied field. The measured Hall angle as a function of temperature for all four samples is presented in Fig.\,\ref{fig:ReImHA}a. The Hall mass extracted from the complex Hall angle is shown in Fig.\,\ref{fig:mHandModelDiagram}. 

\section{Comparison of Low temperature IR Hall results and ARPES}

It is interesting to compare dc and IR magneto-transport data directly with the reconstructed FS properties measured by ARPES. Both dc and IR Hall measurements are sensitive to Fermi surface geometry as well as quasiparticle scattering rates and velocities, precisely the quantities measured by ARPES. We will show that considerable deviations between the information garnered from ARPES data analyzed in terms of the RTA and magneto-transport measurements illustrate a fundamental breakdown of the RTA. Within a Fermi liquid based theory, this breakdown signals a need to properly account for vertex corrections.

Within the RTA in Boltzmann theory, the off-diagonal and diagonal conductivities ($\sigma_{xy}$ and $\sigma_{xx}$) may be expressed as path integrals around the FS involving the Fermi velocity, scattering rate, and size and shape of the FS:\cite{DrewMillisSigmaXX}
\begin{align}\label{eq:RTA}
\sigma_{xy}&=\frac{e^2}{h c_0}\frac{e \vec{B}}{\hbar c}  \cdot
\oint_{\text{\tiny{FS}}} dk \frac{\vec{v}_k^* \times
d\vec{v}_k^*/dk
}{(\gamma^*_k - i \omega)^2} \notag\\
\sigma_{xx}&=\frac{e^2}{h c_0} \oint_{\text{\tiny{FS}}} dk
\frac{|\vec{v}_k^*|}{\gamma^*_k - i \omega}
\end{align}
where $\gamma^*_k= v^*_k \, \Delta k$, $\Delta k$ is the momentum distribution curve (MDC) width as measured by ARPES, $l=1/\Delta k$ is the mean free path, $v^*_k$ is the (renormalized) Fermi velocities measured by ARPES, and $c_0$ is the interplane spacing. In a comparison of ARPES and transport data it is important to recognize that the current relaxation time that characterizes transport is not necessarily the same as the quasiparticle lifetime measured by ARPES.  For example, small angle elastic scattering as seen by ARPES does not affect transport.\cite{DrewMillisSigmaXX}  Therefore in implementing our model we compare the $\gamma^*$ needed to reproduce the transport data to the ARPES values. 

As depicted in Fig.\,\ref{fig:ReImHA}b, we construct a two-arc FS model of the electron pocket centered at ($\pi$,0) whose shape and size is based on ARPES. The corner  and edge arcs as defined in the caption are assigned ARPES measured values of Fermi velocities $v^*_k$ and scattering rates $\gamma^*_k$ computed from the MDC widths. Setting Fermi velocities and scattering rates along each arc to a constant is not strictly correct. If they are considered as effective average values along each arc, then the model captures the quantitative features of the FS properties. Since the discrepancies between ARPES and transport data will be shown to be very large in comparison, errors which may be introduced into the analysis from this approximation are negligible. Utilizing Eq.\,(\ref{eq:RTA}), the complex IR Hall angle and dc Hall coefficient $R_H=\sigma_{xy}/\sigma_{xx}^2$ are calculated and compared to measurements. The results are presented in Table\,\ref{tab:ModelResults}.

\newcommand\T{\rule{0pt}{2.6ex}}
\newcommand\B{\rule[-1.2ex]{0pt}{0pt}}

\begin{table*}[t]
\begin{ruledtabular}
\begin{tabular}{c|ccc|cccc|cc|cccc}
\multicolumn{1}{c|}{Doping} &\multicolumn{3}{c|}{FS Size and
Shape} &\multicolumn{4}{c|}{FS Properties}
&\multicolumn{2}{c|}{Mean Free Path}
&\multicolumn{4}{c}{Modelled/Measured Values}\\[1.5pt]

x (\%) &$k_e$ (1/\AA) &$\beta_e$ &$\kappa$ &$v_e$ (eV \AA) &$\gamma_e$
(eV) &$v_c$ (eV \AA) &$\gamma_c$ (eV) &$l_e$ (\AA) &$l_c$ (\AA)
&$R_H$ &Re($\theta_H$) &Im($\theta_H$) &$m_H$
\\ \hline

10\T   &0.207   &2.50    &0.75    &1.5    &0.06    &0.6    &0.06    &25    &10    &0.4    &0.3    &0.5    &5.6\\
10   &0.207   &2.50    &0.75    &1.5    &0.06    &0.6    &0.02    &25    &36    &1.0    &0.5    &4.2    &8.2\\
10\B   &0.207   &2.50    &0.75    &1.7    &0.06    &3.5    &0.09    &27    &40    &1.0    &1.0    &1.0    &1.0\\
\hline
12\T   &0.221   &7.50    &0.65    &1.5    &0.06    &0.6    &0.06    &25    &10    &0.2    &0.2    &0.1    &2.7\\
12   &0.221   &7.50    &0.65    &1.5   &0.06    &0.6    &0.01    &25    &41    &1.0    &0.7    &2.0    &2.3\\
12   &0.221   &7.50    &0.65    &1.5    &0.07    &1.2    &0.04    &21    &35    &1.0    &1.0    &1.0    &1.0
\end{tabular}
\end{ruledtabular}
\caption{\label{tab:ModelResults}ARPES measured properties of the FS are used to calculate dc $R_H$ and IR Hall angle values under the assumptions of the RTA. Strong deviations are observed (rows 1 and 4). The three parameters used to fit the FS size and shape to ARPES measurements are defined according to the model presented in Fig. \ref{fig:ReImHA}b and associated caption. 
Measured values of dc $R_H$ at low temperature are $-6.5 \times 10^{-9} \Omega \, m$ for the 10\% sample
  \cite{Onose_RH_Optical} and $-6.2 \times 10^{-9} \Omega \, m$ for the 12\% sample.\cite{QCP_16percent}
  Fixing the Fermi velocities to those measured by ARPES and allowing the scattering rates to vary,
  a large mean free path along the corner direction is required to fit the dc $R_H$ (rows 2 and 5).
  Simultaneously fitting the IR Hall data and dc $R_H$ requires a large $v_c$ and $l_c \sim 1.5 l_e$ (rows 3 and 6)
  in contrast to the ARPES measured $l_c \sim 0.4 l_e$ and relatively small $v_c$, illustrating a fundamental breakdown of the relaxation time approximation.}
\end{table*}


Very little ARPES data actually exists for PCCO.\cite{ARPES_PCCO_Richard} However, ARPES measurements have characterized various similar n-type cuprates at a variety of doping levels including Nd$_{2-x}$Ce$_x$CuO$_{4\pm\delta}$,\cite{ARPES_edoped_Matsui, ARPES_edoped_Armitage,ARPES_Schmitt} Pr$_{0.89}$LaCe$_{0.11}$CuO$_4$,\cite{ARPES_Matsui_opt-PLCCO} and Sm$_{1.86}$Ce$_{0.14}$CuO$_{4}$.\cite{ARPES_edoped_Park} 
ARPES measurements together with dc $R_H$ data show fractionalization of a large hole-like FS upon underdoping below a critical value $\sim$16\% into arcs gapped along the magnetic Brillouin zone (BZ).\cite{QCP_16percent,Patrick_New_RH_Data, Lin_Millis,ARPES_edoped_Matsui, ARPES_edoped_Park} Below a doping of $\sim$13\%, the arc in the vicinity of ($\pi/2$,$\pi/2$) is completely gapped out \cite{ARPES_edoped_Matsui} leaving only the arc near ($\pi$,0). In this paper we take the point of view that these observations correspond to the formation of a single convex electron pocket centered at ($\pi$,0) in PCCO for x$<$13\% and that the band-folded sections of FS are much reduced in intensity due to coherence factors in ARPES.\cite{ARPES_edoped_Park}

The FS curvature associated with the edge sections of the electron pocket (defined in Fig.\,\ref{fig:ReImHA}b) is an important consideration when calculating Hall transport. In the case of NCCO, the edge sections have a convex curvature at $\sim$13\%, become straight at around $\sim$15\%, and eventually become concave (or hole-like) at $\sim$16\%.\cite{ARPES_edoped_Matsui} Substitution of various lanthanide elements show a systematic correlation of edge section curvature with atomic number.\cite{ARPES_edoped_Ikeda} The edge sections become hole-like at a cross-over doping level which decreases with a decrease in lanthanide atomic number.\cite{ARPES_edoped_Ikeda} Based upon the measured trends, the curvature of the edge sections at $\sim$12\% of PCCO should be a bit flatter along the edge section than a sample of NCCO at the same doping. For PCCO, we expect a fully convex electron pocket with no hole-like contributions at and below $\sim$12\% doping. Increasing the doping above $\sim$13\%, any hole-like contributions caused from the edge curvature would be more prevalent in PCCO than NCCO.

From the published ARPES data,\cite{ARPES_PCCO_Richard,ARPES_edoped_Matsui,ARPES_edoped_Armitage,ARPES_Schmitt, ARPES_Matsui_opt-PLCCO,ARPES_edoped_Park,ARPES_edoped_Ikeda,PrivateComm} reasonable values of the size and shape of the FS as well as the edge and corner Fermi velocities and scattering rates for excitation energies near the FS (below the observed kink in the energy dispersion)\cite{ARPES_PCCO_Richard} are summarized in Table\,\ref{tab:ModelResults} for 10\% and 12\% doping levels. It should be noted that only one study reports the corner Fermi velocity where $v_c\,\sim\, 0.4 v_e$.\cite{ARPES_edoped_Park} 


We focus upon the two lowest doped samples, 10\% and 12\%, where no hole-like contributions exist either from the hole-pocket near ($\pi / 2$, $\pi / 2$) observed at higher doping or from concave curvature of the edge sections associated with the electron pocket. However, it should be emphasized that our fundamental observation is a very large Hall angle response at low doping expressed as a small Hall mass. Admixtures of hole-like contributions tend to reduce $\sigma_{xy}$ which suppress the Hall response manifesting as a \textit{large} apparent Hall mass. This enhancement of the apparent Hall mass is observed in the 15\% sample, for example (see Figs.\,\ref{fig:ReImHA}a and \ref{fig:mHandModelDiagram}).

The number density calculated from the FS volume measured by ARPES is consistent with the stoichiometric doping to within 10\%. However, the calculated dc $R_H$ \cite{Patrick_New_RH_Data,QCP_16percent} and the IR Hall data for the two lowest doped samples are off significantly compared to the measured values (see Table\,\ref{tab:ModelResults}, rows 1 and 4). It should be emphasized the discrepancy lies well outside of any measurement errors associated with input model parameters or transport values. To fully illustrate this large discrepancy, we consider two cases.

In the first case, we only consider the dc $R_H$ data. The Fermi velocities measured by ARPES are fixed but the scattering rate along the corner sections is allowed to vary in order to fit the dc $R_H$. The result that $l_c \sim 1.5 \, l_e$ (Table\,\ref{tab:ModelResults}, rows 2 and 5) is contrary to ARPES measurements reporting $l_c \lesssim (1/2) \, l_e$.\cite{ARPES_edoped_Park}

A similar comparison between the low temperature dc $R_H$ data and values calculated from a mean field SDW model was performed by J. Lin and A. J. Millis.\cite{Lin_Millis} They used a tight binding band model with a doping dependent SDW gap which produces a Fermi surface whose size and shape is very similar to that measured by ARPES.  The Fermi velocities were weakly anisotropic such that $v_c/v_e\sim0.8$, compared with $v_c/v_e\sim0.4$ measured by ARPES. An isotropic scattering rate was assumed. Our model reproduces the same calculated Hall coefficient using the Fermi velocities calculated from their tight binding model. The trend observed in their study, that the measured dc $R_H$ is significantly larger than expected, becomes much more pronounced when the actual anisotropic velocities as measured by ARPES are used to calculate transport values.

Although the dc $R_H$ can be fit by only adjusting the corner scattering rates, the IR Hall data is off by over a factor of 2 and cannot be fit under any parameter values assuming the ARPES measured Fermi velocities. Therefore, in the second case, we allow the Fermi velocities as well as scattering rates to vary to fit both the dc $R_H$ as well as the low temperature IR Hall data. The resulting values are significantly different from those measured by ARPES (Table\,\ref{tab:ModelResults}, rows 3 and 6). We still require $l_c \sim 1.5\, l_e$ in order to fit the dc $R_H$, but to concurrently fit the IR Hall data requires a large $v_c$ whereas ARPES measures a relatively small $v_c$. 

This large discrepancy can be intuitively understood. For a simple one-band isotropic FS, $m_H \sim k_f/v_F$, so extremely small Hall masses result from strongly curved sections of FS (a small radius of curvature $k_F$, like at the corner sections of the electron pocket) and a large Fermi velocity. The more rigorous model shows this precise behavior requiring large velocities along the corner sections of FS (Table\,\ref{tab:ModelResults}, rows 3 and 6). The large discrepancy between the analyzed ARPES data and the dc and IR magneto-transport data elucidates a significant breakdown of the RTA approximation.

There are other observations which cast doubt on a simple SDW model fully describing the ground state in the underdoped n-type cuprates. The large increase in $v_c$ required to describe the IR Hall data when doping from 12\% to 10\% contradicts the mean-field SDW prediction. Using reasonable values in a tight binding model (t, t', t'', and changes of the gap energy and Fermi energy with doping), a very gradual decrease in the velocities $\lesssim$25\% is expected as the doping is changed from 12\% to 10\%. However, in order to describe the IR Hall data, an increase in $v_c$ by a factor of $\sim$3 is required. Furthermore, the same tight binding model predicts very little anisotropy of the Fermi velocity associated with the electron pocket such that $0.7 \lesssim v_c/v_e \lesssim 0.8$. However, the measured anisotropy by ARPES is a much larger factor, $v_c/v_e \approx 0.4$. Note that to fit the IR Hall data tabulated in Table\,\ref{tab:ModelResults} (rows 3 and 6) requires anisotropies in velocity in the opposite sense, $v_c/v_e \approx 2$ for the 10\% sample. Lastly, ARPES data shows a weak remnant large unreconstructed hole-like FS superimposed upon the reconstructed FS suggesting that a simple SDW may not capture all the physics associated with the underdoped state.\cite{ARPES_edoped_Park} However, it should be noted that this effect would tend to suppress the Hall angle response due to the admixture of hole-like contributions reducing $\sigma_{xy}$, a trend which is contrary to the observed large dc and IR Hall response.

In this section we have shown that the low temperature dc $R_H$ response associated with the 10\% and 12\% doped samples is anomalously large compared to that expected from ARPES measurements. Our measured IR Hall angle response was also shown to be anomalously large, characterized by a small and doping dependent Hall mass. These deviations were shown to lie well outside the error bars associated with any of the measurements.  The discrepancy between ARPES and magneto-transport measurements signals a breakdown of the RTA.

\section{Temperature dependence of the Hall mass}

In Fig.\,\ref{fig:mHandModelDiagram}a, the three lowest doped samples show a gradual increase in Hall mass parameter with temperature. This is qualitatively consistent with a rollover from an electron-like FS to a large hole-like FS.  In this scenario, the Hall angle would be expected to be negative at low temperatures eventually becoming positive, crossing zero in the interim. At the zero crossing, the apparent Hall frequency would be zero and the corresponding Hall mass parameter would be infinite. The gradual rise of the Hall mass parameter observed in Fig.\,\ref{fig:mHandModelDiagram}a corresponds to a fractionalized FS slowly recovering to a large hole-like response at high temperature. A similar behavior is observed in the dc $R_H$ \cite{QCP_16percent, Patrick_New_RH_Data, Onose_RH_Optical} and dc $\cot(\theta_H)$ \cite{Dagan_RHvsDoping_lowTemp} for similarly underdoped samples in which the response is electron-like at low temperature and gradually decreases in magnitude with increasing temperature.

However, the N\'eel temperature for the n-type cuprates is relatively low as depicted in Fig.\,\ref{fig:mHandModelDiagram}a: $T_N|_{10\%}\,\approx\,150$K, $T_N|_{12\%}\,\approx\,100$K, and $T_N|_{13.5\%}\,\approx\,0$.\cite{Motoyama} Well above the N\'eel temperature where antiferromagnetic (AFM) fluctuations are negligible we expect a recovery of a large hole-like FS. A continuing electron-like response with a gradually changing Hall mass is observed well above the N\'eel temperature, an observation consistent with dc $R_H$ measurements.\cite{Onose_RH_Optical,QCP_16percent, Dagan_RHvsDoping_lowTemp} We interpret this as evidence of current vertex corrections induced by AFM fluctuations in the paramagnetic state. In overdoped PCCO, the current-vertex corrections were shown to successfully describe the salient features of the dc and IR Hall angle measurements as a function of temperature, doping, and frequency.\cite{Jenkins_Overdoped}

In a simple SDW scenario, a temperature independent Hall mass is expected well below the N\'eel temperature. Together with the observation that the ARPES measured FS properties strongly deviate from dc and IR magneto-transport data, the clear temperature dependence of the Hall mass well below the N\'eel temperature we interpret as additional evidence that a simple SDW model is not a viable description of the ground state of the system. 

\section{Discussion}

In this section we discuss the possible origins of the observed discrepancies of the dc and IR magneto-transport data with ARPES data analyzed within the RTA. First we note that the discrepancy associated with the dc $R_H$ is not explained by Mott correlations since $R_H$ is not affected by the quasiparticle spectral weight. This suggests that vertex corrections to the conductivity are the relevant effect. Vertex corrections arising from AFM fluctuations in the paramagnetic state account for the anomalous dc and IR magneto-transport in overdoped PCCO.\cite{ Kontani-ACHall1, Kontani_2008Review,Jenkins_Overdoped}  

Further insight into the nature of the ac magneto-transport can be obtained from the following expression for the Hall mass:  
\begin{eqnarray}
\frac{m_H}{m_e}  = -\frac{\omega_c^e}{\omega}\frac{\left|\sigma_{xx}\right|}{\left|\sigma_{xy}\right|}\sin(\phi_H - \phi_C)\label{eq:mHphase}
\end{eqnarray}
where $\sigma_{xx}=|\sigma_{xx}|e^{i\phi_C}$, $\sigma_{xy}=-|\sigma_{xy}|e^{i\phi_H}$, and $\omega_c^e = 0.115 meV / T$ is the bare electron cyclotron
frequency.

From Eq.\,\ref{eq:mHphase}, it can be seen that a reduced $m_H$ can occur in two ways.   First, $|\sigma_{xx}|$ can reduce more rapidly than $|\sigma_{xy}|$.  This is the expectation for small pockets as $\sigma_{xx}$ is proportional to the Fermi circumference and $\sigma_{xy}$ tends to be less dependent on the size of the Fermi pocket (see Eq.\,\ref{eq:RTA}).  Note that a mixture of hole-like and electron-like contributions to $\sigma_{xy}$ leads to a reduction in $|\sigma_{xy}|$ and a corresponding \textit{increase} in $m_H$. Alternatively a strong reduction of $m_H$ can occur due to a diminishing phase difference.  Moreover, to achieve a zero crossing of $m_H$ as suggested by the data in Fig.\,\ref{fig:mHandModelDiagram}b below a doping of $\sim9$\% it is necessary to have a sign change of $\phi_H-\phi_C$.

Therefore the observed rapid reduction in $m_H$ may signal a reduction in $\phi_H$ relative to $\phi_C$.  Some insight into this possibility is gained by expressing the conductivity in an extended Drude parameterization.  We can write to lowest order in $\omega/\Gamma$:
\begin{eqnarray}
\phi_H  \cong \frac{2 \omega (1+\lambda_H)}{\Gamma_H} \text{\,\,\,\,\,\,\,and\,\,\,\,\,\,\,} \phi_C  \cong \frac{\omega (1+\lambda_C)}{\Gamma_C} \label{eq:mHphaseExtDrude}
\end{eqnarray}
Where $\Gamma_H$ ($\lambda_H$) and  $\Gamma_C$ ($\lambda_C$) are the relaxation rates (mass enhancement parameters) for $\sigma_{xy}$ and $\sigma_{xx}$ respectively.  From Eqs.\,\ref{eq:mHphase} and \ref{eq:mHphaseExtDrude} it is seen that a reduced $\lambda_H$ with a nearly constant $\Gamma_H$ will lead to a reduced $m_H$.  The $\lambda_H$ / $\lambda_C <1$ behavior is expected based on frequency dependent IR Hall studies of BSCCO and YBCO.\footnote{In mid IR Hall measurements a reduced  $\Gamma_H$ compared with $\Gamma_C$ is observed.  By Kramers-Kronig relations this implies that $\lambda_H$ / $\lambda_C <1$ in the far IR spectral region.}$^,$\cite{SchmadelPRBRapid} These effects can arise from vertex corrections.\cite{Kontani-ACHall1, Kontani_2008Review,SchmadelPRBRapid} 

In underdoped PCCO the vertex corrections from magnetic fluctuations should be significantly modified from those observed in overdoped PCCO due to the long range magnetic order.\cite{Jenkins_Overdoped}  That is, below the N\'eel temperature, AFM fluctuations are absent so that vertex corrections would need to arise from some other interaction effect. Apart from phonons, the possibilities are charge and magnons fluctuations.   

In the magnetically ordered phase the magnon excitations near the center of the magnetic Brillouin zone (BZ) can mediate the electron-electron interaction in the small electron pocket. The strong electron-magnon interaction in the cuprates implies that these interactions can affect the electron transport. However, the discrepancies of the experimental data with the RTA means that vertex corrections are needed to explain the data in terms of the electron-magnon interaction.  

The observed temperature dependence of $R_H$ and $m_H$ support this scenario. $R_H$ and $m_H$ in the paramagnetic phase are electron-like despite the expected hole-like FS in the absence of the SDW gap.  The negative $R_H$ and $m_H$ in the paramagnetic phase can be explained by vertex corrections due to the electron-electron interactions induced by magnetic fluctuation similar to that observed in overdoped PCCO.\cite{Jenkins_Overdoped} Moreover, the absence of a magneto-transport signature of the N\'eel temperature indicates a continuous transition from magnetic fluctuation induced CVC in the paramagnetic phase to magnon induced CVC in the antiferromagnetic phase. 

The striking observation of a rapid reduction of $m_H$ with underdoping appears to imply a rapid decrease in the phase of $\sigma_{xy}$ compared with that of $\sigma_{xx}$ and may provide an important clue as to the nature of the corresponding interaction induced corrections to RTA.     

Charge fluctuations related to stripe order can also lead to vertex corrections.  The nearly square electron Fermi surfaces observed in ARPES implies nesting associated with the nearly flat edges. This nesting could lead either to a charge density wave ground state with long range charge order or vertex corrections to the conductivity due to charge order fluctuations. These two possibilities produce similar effects, namely a reduction of the low frequency contributions to $\sigma_{xx}$ from some of the edge segments of the FS.  A reduction in $\sigma_{xx}$ implies an enhancement of $R_H$ and $\theta_H$ (and therefore a suppression in $m_H$). However, CDW ordering would be expected to reconstruct the electron pocket leading to nearly one dimensional segments of FS with some hole-like curvature induced by the CDW gap thus reducing $\sigma_{xy}$ as well. Furthermore, the electron pocket associated with the 12\% sample is more square than that of the 10\% sample. Any CDW effects observed in transport measurements should be more pronounced in the 12\% pocket although the Hall mass continues to dramatically decrease with underdoping. While we believe these arguements make the CDW scenerio unlikely a theoretical analysis of the mixed SDW and CDW effects on the transport would better test this possibility.

\section{Conclusion}
We measured the IR Hall angle as a function of temperature and doping at an excitation energy of 10.5 meV. Although the data is qualitatively consistent with the formation of a small electron pocket in underdoped PCCO, neither the dc $R_H$ data nor IR Hall angle response at low doping is consistent with ARPES data when analyzed under the assumptions of the relaxation time approximation. The large discrepancy between transport and ARPES data demonstrates a clear breakdown of the RTA implying the importance of including vertex corrections in the conductivity.  A negative and continuously increasing Hall mass well above the N\'eel temperature observed in all underdoped samples is difficult to explain within a SDW model in the ordered phase, but can be understood in terms of current vertex corrections from AFM fluctuations in the paramagnetic state, similar to the case of overdoped PCCO.\cite{Jenkins_Overdoped}

\acknowledgements
The authors wish to acknowledge the support of NSERC, FQRNT, CFI, CNAM and NSF (DMR-0653535 and DMR-
0303112). We wish to thank N. P. Armitage, M. Ikeda, V. Galitski, H. Kontani, T. D. Stanescu, and T. Takahashi and  for insightful discussions. We thank S. Pelletier and Dr. K. D. Truong for their technical assistance.

\bibliography{Underdoped_PCCO}

\end{document}